\documentclass[traditabstract]{aa}

\usepackage{graphicx}
\usepackage{natbib}

\date{To appear in the A\&A special issue for GREAT/SOFIA}

\titlerunning{GREAT/SOFIA atmospheric calibration}
\authorrunning{Guan et al.}

\begin{document}


\title{GREAT/SOFIA atmospheric calibration}


\author {
    Xin Guan\inst{1},
    J\"urgen Stutzki\inst{1},
    Urs U. Graf\inst{1},
    Rolf G\"usten\inst{2},
    Yoko Okada\inst{1},
    Miguel Angel Requena Torres\inst{2},
    Robert Simon\inst{1},
    Helmut Wiesemeyer\inst{2},
}


\institute{I. Physikalisches Institut, Universit\"at zu K\"oln,
Z\"ulpicher Str. 77, K\"oln, Germany, 50937
\and
Max-Plank-Institut f\"ur Radioastronomie,
Auf dem H\"ugel 69, Bonn, Germany, 53121}


\abstract{

  The GREAT observations need frequency-selective calibration across
  the passband for the residual atmospheric opacity at flight
  altitude. At these altitudes the
  atmospheric opacity has both narrow and broad spectral features.
  To determine the atmospheric transmission at high spectral resolution,
  GREAT compares the observed atmospheric emission with atmospheric
  model predictions, and therefore depends on the validity of the atmospheric
  models. We discusse the problems identified in this
  comparison with respect to the observed data and the models, and
  describe the strategy used to calibrate the science data from
  GREAT/SOFIA during the first observing periods.

}


\maketitle

\section{Introduction}

The {\it German REceiver for Astronomy at Terahertz frequencies}
(GREAT)\footnote{GREAT is a development by the MPI f\"ur
Radioastronomie (Principal Investigator: R. G\"usten) and
KOSMA$/$Universit\"at zu K\"oln, in cooperation with the MPI f\"ur
Sonnensystemforschung and the DLR Institut f\"ur Planetenforschung.}
\citep{hey12} onboard the Stratospheric Observatory for
Infrared Astronomy (SOFIA) \citep{bec09,you12} observes at typical flight
altitudes of 8 km (where telescope- and instrument set-up starts)
up to 14 km.
The atmospheric transparency at these altitudes is high
(though not unity) except near strong absorption lines of species such
as water and other trace constituents 
(${\rm O}_3$, ${\rm CO}$, ${\rm N}_2{\rm O}$, etc.).
The typical amount of water vapor in the atmosphere above
the observatory (precipitable water vapor, $pwv$)
is in the range of a
few $\mu$m at very good observing conditions and up to 50 
$\mu$m and higher at bad conditions. The atmospheric pressure, rapidly
decreasing with increasing altitude, results in narrower absorption
lines with increasing altitude. As a result, the atmospheric opacity at
flight altitude shows many narrow spectral features, typically with a
width of about 100 MHz down to a few MHz (varying with species and
altitude at which the particular species absorbs). In addition,
extended line wings from very strong absorption lines, in particular
of water, and collision-induced absorption (CIA), where the
collisions of the absorbing species with other molecules generate a
temporary dipole moment, contribute to a spectrally
smooth broad-band opacity.

The correction of airborne astronomical data for the atmospheric
transmission is thus much more challenging than for ground-based
observations, because of the small-scale variations resulting
from the narrow atmospheric lines. Direct measurement of the
frequency-dependent transmission is in principle possible, but is
limited by the inherently higher noise in narrow frequency-bins.
Hence, the calibration has to rely on appropriate models of the
frequency-dependent atmospheric opacity and the proper approach is to
determine the transmission from model fits to the observed sky
emission. This approach has been implemented and successfully used in
recent years at ground-based high-altitude observatories such as
APEX \citep{gues06} and NANTEN2 \citep{kaw05} in Chile.
The proper description of the frequency-dependent
atmospheric transmission is particularly important for
absorption measurements against continuum background sources, because the
background emission is also modulated by the atmospheric transmission,
which can therefore interfere with the absorption profiles from the
astronomical sources.


We first discuss the calibration approach in
section~\ref{sec:approach} and its implementation for
GREAT on SOFIA in section~\ref{sec:GREAT_impl}.
GREAT allows us for the first time to
simultaneously observe two widely separated frequency bands in the THz-regime
at high spectral resolution. Accordingly, it has come as no
surprise that the atmospheric models currently in use for
predicting the THz-regime emission and absorption of the atmosphere turn
out to have clear deficiencies in properly modeling the atmosphere
above SOFIA to the degree necessary to consistently calibrate the
GREAT data. These problems are discussed in section
\ref{sec:GREAT_data}. We
conclude by describing the pragmatic approach chosen for the
calibration of the GREAT data of the first observing season and the
perspectives for future improvements in section \ref{sec:summary}.

\section {Approach to atmospheric calibration} \label{sec:approach}

Following the notation of Appendix~\ref{app:calib}, the sky brightness
on an equivalent antenna temperature scale along a line-of-sight at
elevation $El$ for a plane parallel atmosphere,
i.e.\ taking for the airmass the approximation $A=1/\sin(El)$, is given by
\begin{eqnarray}
T_{A,sky,\nu}
&=&
\int_0^{\tau_{Z,\nu}}
{\cal J}_\nu\left(T
({\tau}_\nu ')\right) \times 
\,e^{ 
-{\tau}_\nu ' \,A}
\,
d\left({\tau}_\nu ' \,A \right)\\
&=:&
{\tilde T}_{sky,\nu}\left(1-t_{a,\nu}\right),
\end{eqnarray}
with the atmospheric transmission
\begin{equation} 
t_{a,\nu} = \exp\left( -\tau_{Z,\nu}/\sin(El)\right),
\end{equation}
the frequency-dependent zenith opacity $\tau_{Z,\nu}$,
and the definition of the effective sky brightness 
\begin{equation}
\label{equ:app-Ta_main}
{\tilde T}_{sky,\nu}=T_{A,sky,\nu}/\left( 1-t_{a,\nu}\right)
\end{equation}
according to equ.~\ref{equ:app-Ta}.

An often used approximation is that the emission occurs at the ambient
temperature of the atmosphere near the observatory, $T_{amb}$, so that
\begin{equation}
{\tilde  T}_{sky,\nu}\approx {\cal J}_\nu(T_{amb}),
\end{equation} 
although this assumption may
not be particularly well justified for absorption lines from
the higher atmosphere, which is warmer than at 
flight altitude (see Section~\ref{sec:GREAT_impl}).

In principle, the above equation can be inverted to calculate the sky
transmission at each frequency from the measured sky brightness. 
However, because the atmospheric calibration has
to be frequency specific, deriving the opacity from the
sky-emission in each frequency bin results in
quite large errors due to the limited signal-to-noise ratio in their narrow
band-width. In addition, a heterodyne instrument
receiving double-sideband (such as GREAT)
folds the signal- and image-sideband on top of each other, so
that the
emission from the two sidebands cannot be
separated. To overcome these problems, one has to (i) use an atmospheric
model to predict the atmospheric opacity as a function of frequency at
the given observing altitude and line-of-sight, parameterized by a few
dominant atmospheric parameters, such as the precipitable water vapor and
 the ambient temperature, (ii) derive these parameters from the
comparison between the observed emission and the model, and (iii) then use
the model opacities, calculated with these best-matching parameters,
to correct for the attenuation of the astronomical signal.

This can be done either for individual spectrometer channels to only
de-convolve the image- and signal-sideband opacity, or, to increase
the signal-to-noise ratio, over selected sub-bands or the
complete reception bandwidth of the receiver (which in the case of
GREAT might include the two simultaneously observed THz-bands
L1 and L2). A necessary condition for the consistency of the
calibration approach is that the atmospheric parameter(s) derived for
individual channels or sub-bands are identical to within the noise
and/or that the residual of the comparison between the atmospheric
model and the observed sky-brightness is within the observed noise
level. If this is not met, the atmospheric model obviously does not
describe the actual atmosphere satisfactorily
(or the measurements are corrupt).

The zenith opacity above the observer at altitude $h$ 
depends on the altitude profile of the abundances of the atmospheric
species that are responsible for the absorption, which is driven by the
complex chemical processes in the atmosphere, as well as by the pressure
profile (influencing the pressure broadening) and temperature profile
(determining the population of the relevant species within the
absorbing states).  All these may vary as a function of geographical
position and local weather parameters. To arrive at a
manageable problem, one has to make additional assumptions, such as
that the altitude variations of abundances, temperatures, and pressure
are given by a {\it standard atmosphere}, e.g., the \citet{us76},
which specifies the altitude variation on a {\it pressure altitude}
vertical scale (in the following, {\it altitude} often refers to {\it
  pressure altitude}).  Similarly, it specifies the total abundances
and abundance variation with altitude of the atmospheric species of
interest, except for a few, where their distribution has to be adapted
to the specific observing conditions, as discussed below. In
particular the column of water molecules above the observer,
$pwv$, may change rapidly, because it is largely determined by local
weather. Hence, a useful parameterization of the atmospheric
absorption is given by splitting the {\it wet}, i.e., water absorption,
and the {\it dry} part as
\begin{eqnarray}\label{equ:bc}
\tau_{0,\nu}&=&b_\nu(p_{amb}[,T_{amb},X_j,\ldots]) \times pwv+\nonumber\\
&&c_\nu(p_{amb}[,T_{amb},X_j,\ldots]),
\end{eqnarray}
where the wet and dry coefficients are a function of the ambient
pressure $p_{amb}$ (or altitude), and additional atmospheric
parameters such as the ambient temperature $T_{amb}$, abundances $X_j$,
and possibly other parameters that, e.g., specify the altitude profile
of these abundances. A particular atmospheric model can then be
represented by a multi-dimensional look-up-table of the $b,\,c$
-coefficients vs. frequency, (pressure) altitude, and additional
parameters, which are derived from the various atmospheric models by
fitting the model opacities, calculated at two values of $pwv$ and
at different frequencies and
altitudes. The details on deriving the $b$- and $c$-coefficients are
given in Appendix~\ref{app:calib}.

\subsection{Atmospheric models}
Several atmospheric
models are in use within the radio- and FIR-astronomy community, both for
ground-based observatories and air-borne applications.  For the
GREAT/SOFIA calibration we have investigated the use of
AM \citep{am},
ATRAN \citep{atran}, and MOLIERE \citep{moliere}. The ATM-model
\citep{atm}, in use for ALMA and other ground-based observatories, in
its publically available form does not include the higher excitation
water lines and hence is of limited use for frequencies above 1.5~THz.

\begin{figure*}
\centering
\includegraphics[width=\textwidth]{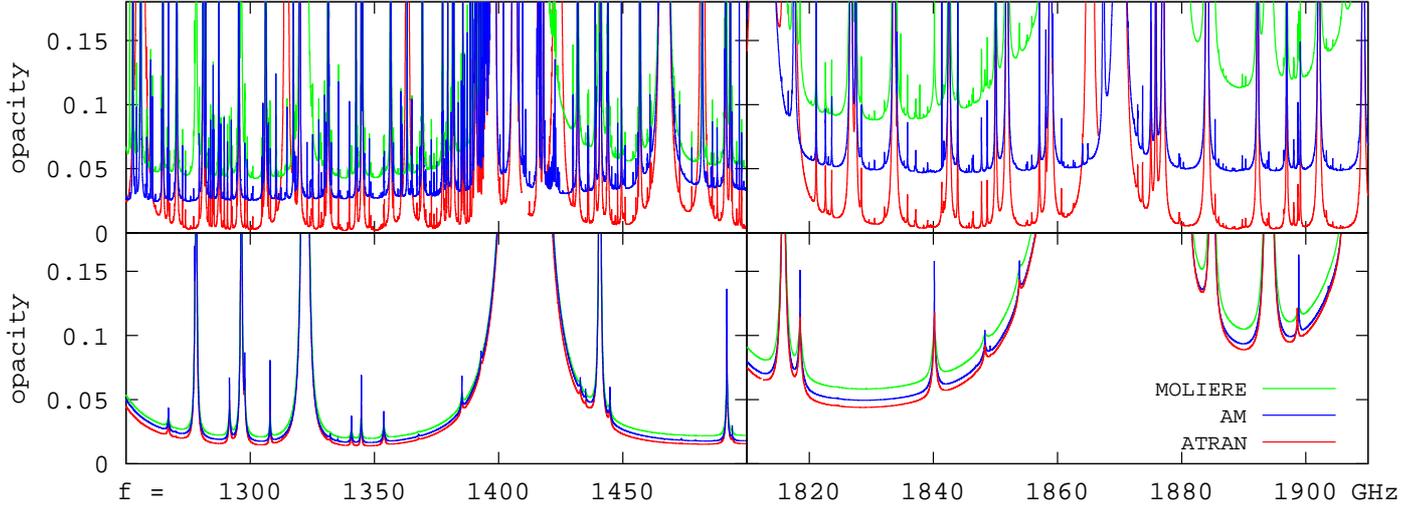}
\caption{Comparison of the dry (top) and wet (bottom) opacity
  coefficients as defined in equ.\ \ref{equ:bc}, \ref{equ:b_h} and
  \ref{equ:c_h} for the frequency range of GREAT channels L1 (left) and
  L2 (right), calculated from the atmospheric models AM, ATRAN and
  MOLIERE (see text), showing the opacity for a pressure altitude of
  $p_{amb}=133 {\rm Torr}$ and a $pwv=10 {\rm \mu m}$. The largest
  discrepancy is in the quasi-continuous dry-atmospheric opacity due
  to collision- induced absorption (CIA), which is completely neglected
  in ATRAN and differs by about a factor of 2 between MOLIERE and
  AM.}
\label{fig:atm_overlay}
\end{figure*}

Figure~\ref{fig:atm_overlay}, comparing the {\it wet-} and {\it dry-}
atmospheric coefficients of the three models, shows that ATRAN
at present does not include the collision-induced absorption (CIA)
from $N_2$ and $O_2$, which is responsible for the quasi-continuum opacity
shown in the other two models. The atmospheric opacity due to these
CIA-processes differs by a
factor of $\approx 2$ between MOLIERE and AM, being higher for MOLIERE. 
All three models agree on the {\it wet}
atmospheric opacity, although ATRAN has some small errors in the
frequencies of a few higher excitation water lines.
Because ATRAN does not include the dry-atmosphere continuum opacity, and
the dry-atmosphere contribution to the opacity due to CIA-processes is
already too high compared to the observed sky emission (see below)
even for the AM model, we used the AM model for the standard GREAT
calibration.

To calculate the atmospheric transmission for different
$pwv$-values, we modeled the water mixing ratio as an exponential
function with altitude in the high troposphere and low stratosphere,
and with a constant, but low value higher above,
following \citet{mas68} and \citet{her09}.
The detailed treatment is presented in Appendix~\ref{app:calib}.

The nitrous oxide, N$_{2}$O, profile is known from ground- and balloon
observations \citep{emm94,cam04,str08}. We adapted a simplified
profile for use with AM. It uses a N$_{2}$O mixing ratio of 0.3~ppm
below an altitude of 60~hPa, and 0.1~ppm above.  This is a rough
approximation of many measured profiles, such as that given in
Figure~1 of \citet{cam04}.  N$_{2}$O does not produce particularly
strong atmospheric features, but we adapted an N$_{2}$O profile based on
real observations, to compare AM with ATRAN, which produces
far too strong N$_{2}$O features compared with the observed sky
emission.

The ozone, O$_{3}$, profile changes rapidly with geographic positions
and with season. The profile actually used in the GREAT calibration is
obtained by fitting sky measurements with the NANTEN2 telescope in
Chile to properly reproduce the strength of the ozone lines.  The
GREAT/SOFIA calibration shows that the measured ozone lines are adequately
reproduced by the model, so that this altitude profile is also
applicable here (although a few SOFIA measurements show an increased
ozone abundance compared to this model, which does not affect the
overall calibration, because the ozone lines are very narrow spectral
features, but has to be taken into account for absorption measurements
against a continuum source).
%

\section{Implementation for GREAT}
\label{sec:GREAT_impl}

The calibration of atmospheric transmission for GREAT/SOFIA is
implemented in the task {\it kalibrate} as part of the {\it
  kosma\_software} observing software package. The same scheme is used
for the submm-wave data at the NANTEN2 observatory in Chile.  It fits
the observed, calibrated differential 
sky--hot spectrum with the atmospheric model
brightness as given by equs.~\ref{equ:A_skybrightness} and
\ref{equ:abbrev} (and uses the approximation $\omega=\Omega=0$ and
${\tilde T}_{sky,\nu}={\cal J}_\nu (T_{sky})$ as detailed in
Appendix~\ref{app:calib}). The best-fit value is thus determined with high
precision, making use of the full spectral information observed, i.e.,
with a good signal-to-noise ratio. The user can control whether the fit is
performed as a ``common'' fit to all simultaneously observed spectra
(several receiver pixels at several frequency bands) or individually
for each spectrometer.  The atmospheric transmission for this
best-fit-value is then applied to the full spectra channel by channel,
thus keeping the full frequency resolution of the atmospheric features
(to the resolution with which the frequencies where sampled when generating
the $b$-,$c$-coefficient look-up tables).

\begin{figure}
\centering
\includegraphics[width=0.48\textwidth]{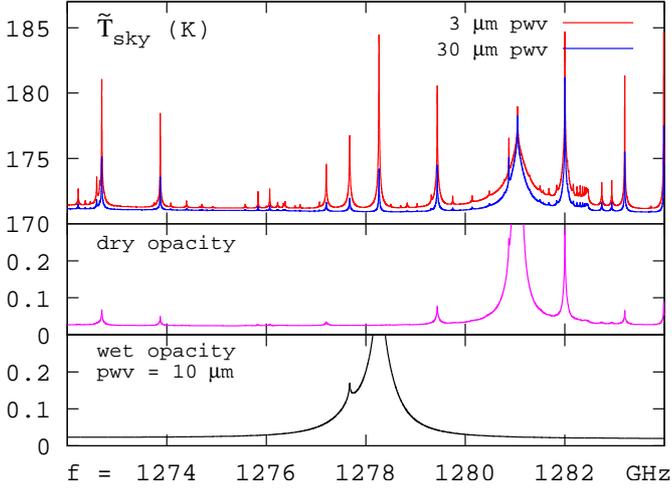}
\caption{ {\it Effective sky brightness}, ${\tilde T}_{sky}$, (top
  panel) in a representative frequency window near a water line and
  several dry-atmospheric lines originating in the higher, warmer
  atmosphere, calculated with the AM model at a pressure altitude of
  133~Torr and an ambient temperature of 200~K, for two values of
  $pwv$, 3 (red) and 30~$\mu$m (blue).  The lower two panels show the
  dry atmospheric opacity (i.e., the $c$-coefficient; middle) and the
  wet atmospheric opacity at 10~$\mu$m $pwv$, i.e.,
  $b\times\,10\,\mu{\rm m}$; bottom), similar to
  Fig.~\ref{fig:atm_overlay} to identify the spectral features
  causing the excess {\it effective brightness temperature}.}
\label{fig:tTsky_bc}
\end{figure}

The approximation ${\tilde T}_{sky,\nu}={\cal J}_\nu (T_{sky})$
assumes that all opacity originates at the same temperature.
To check this assumption, Figure~\ref{fig:tTsky_bc} shows the {\it
  effective sky brightness} ${\tilde T}_{sky,\nu}$ (equ.\
\ref{equ:app-Ta_main}), calculated with the AM model from the model
brightness and opacity for a frequency range near a strong water line
at two different $pwv$-values. The dry-atmospheric lines (shown in the
middle panel in Figure~\ref{fig:tTsky_bc}), as well as the narrow core
of the high excitation water line (see lower panel in
Figure~\ref{fig:tTsky_bc}), originate in the higher stratosphere,
where the temperature increases well above the ambient temperature at the
observer's altitude in the lower stratosphere.  At the low $pwv$ of 3
$\mu{\rm m}$, the effective sky brightness temperature in these lines
is therefore substantially higher than the Rayleigh-Jeans-(R-J)
corrected ambient sky
temperature of about 172~K. With higher $pwv$, the increasing opacity
at altitudes, and hence temperatures, near the observer's altitude
absorbs these emission lines from the high atmosphere, so that the
excess effective sky temperature near these lines is reduced and the
effective sky temperature approaches the ambient temperature.  The
excess is typically up to about 10~K at very low $pwv$. Neglecting it
results in a slight underestimation of the sky brightness in the cores
of these lines for the model fit to the observed brightness.
Because this affects only narrow frequency ranges,
it has a negligible effect on the derived fit parameters.

\begin{figure}
\centering
\includegraphics[scale=0.4,angle=-90]{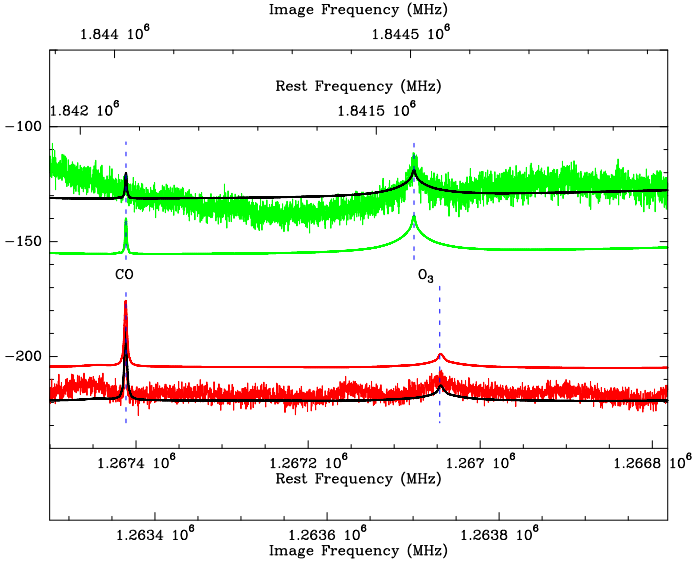}
\caption{
  Example T$_{sky}$~-~T$_{hot}$ spectra from simultaneous
  observations in the L1 (red) and L2 (green) band, showing two ozone
  lines (the broad features toward the middle of the L1 and L2
  passband) and two CO lines (the narrow peaks on the left). Note that
  the two CO lines appear at the same velocity on a $v_{LSR}$ scale,
  because the receiver was tuned to two CO lines for the astronomical
  observations.  The two black curves are independent fits of
  T$_{sky}$~-~T$_{hot}$ for the two bands.  The red and green lines
  between the measured spectra are the result of a {\it common} fit,
  which simultaneously fits a single value of $pwv$ to both bands.
  In a common fit, a consistent value of $pwv$ cannot be derived for
  both bands.  }
\label{fig:common_fit}
\end{figure}

\begin{figure}
\centering
\includegraphics[width=0.48\textwidth]{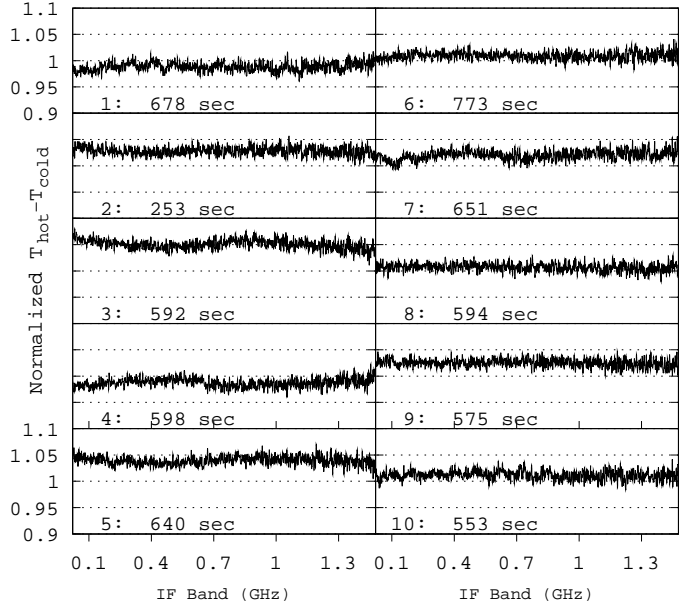}
\caption{ Time series of hot--cold load-calibration scans (in the GREAT
  L2 channel), displayed across the central part of the IF band of
  width 1.4~GHz, calibrated against the previous load scan and
  normalized to the R-J-corrected hot--cold brightness difference. It
  shows that the receiver gain drifts by up to a few percent over time
  scales of several minutes. The sequence number and time offset to
  the previous scan is given in the bottom left corner of each panel.
}
\label{fig:load_cal}
\end{figure}

\section{Application to GREAT data}
\label{sec:GREAT_data}

This calibration scheme, when applied to the GREAT data, is not able
to self-consistently fit the atmospheric emission simultaneously for
the L1 and L2 bands, whereas independent fits to the individual
receiver bands converge well, but on solutions with different
values of $pwv$ for each band. Comparison of the fits with the
observed sky--hot spectra shows that the narrow spectral features
resulting from the dry atmosphere and the line features of water
are reproduced well.  This indicates that the problem lies in the
quasi-continuum opacity of the dry atmosphere, i.e., the approximately
constant offset of the $c$-coefficients visible in
Fig.~\ref{fig:atm_overlay} (resulting largely from
the CIA-processes of N$_2$ and O$_2$) predicted in the models.

A closer look confirms this and shows that this
quasi-continuum opacity of the dry atmosphere is too high.  
Indeed, independent fits to the emission in the two receiver bands for
observations with very good transmission converge to a formal fit
result with slightly negative $pwv$-values: because the model opacity
cannot be reduced below the dry-atmospheric contribution, i.e., the
(too high) value of the $c$-coefficient, this excess opacity has to be
compensated by negative $pwv$, and hence a negative wet opacity, $b
\times pwv$, to match the observed low brightness of the sky
at these low opacities (because negative values for $pwv$ are unphysical,
the fit result is reset to $pwv=0$ in these cases).

Observations at higher overall opacity, i.e., low elevation and higher
$pwv$, show a similar discrepancy.  Figure\ \ref{fig:common_fit}\ shows
such an example 
for a tuning that observes the CO (11--10) line in L1
and the CO (16--15) line in L2.  Separate fits (black lines in
Figure~\ref{fig:common_fit}) give $pwv$=12.3~$\mu$m\ for the L1 band
and 35~$\mu$m for the L2 band.  
The excess opacity in the model 
needs to be compensated in the fitting by a more reduced wet
opacity, hence reduced $pwv$ values, than would be needed with
the proper dry opacity.  The scaling factor between the wet opacity
and $pwv$, namely the $b$-coefficient as defined above, is typically
higher by a factor of 2 to 3 for the L2 band, so that the excess dry
opacities in the model result in accordingly scaled, i.e., smaller differences
in $pwv$ for the fitted $pwv$-value for the L2 band compared to L1.

Similarly, a common fit to both bands, now forced to using a unique
value for $pwv$, results in a $pwv$-value 
somewhere in between the two discrepant values fitted
in the separate fits. It overestimates the opacity in the band with lower
opacity and underestimates the opacity in the other, higher
opacity, band as shown by the green and red fit line in
Figure~\ref{fig:common_fit}, which corresponds to a best-fit value of
21.4~$\mu$m $pwv$. 

The above holds for tunings where the wet atmosphere does not
contribute any line feature, but merely a continuum opacity (with a
slight slope).  In the rare cases
where the wet atmosphere also contributes with a line feature,
even the
single frequency band fits may not converge to a consistent result: the
line intensity forces a certain amount of water and hence also
continuous opacity, so that the total opacity, including the dry
opacity overestimated by the atmospheric model, overshoots the range
that is consistent with the observed atmospheric emission.

To confirm these findings, we modified the fitting procedure to include 
additional fit parameters $o_{L1}$ and $o_{L2}$, namely a
frequency dependent dry-offset, so that equ.\ \ref{equ:bc} changes to
\begin{equation}
\tau_\nu=b_\nu\,pwv + c_\nu -\left\{
\begin{array}{ll}
o_{L1},&\nu\ {\rm in\ L1}\\
o_{L2},&\nu\ {\rm in\ L2}
\end{array} \right. .
\end{equation}
This gives a more consistent fit,
but not with unique offsets $o_\nu$ for different tunings and an
ambiguous separation into $pwv$ offsets and values for $o_\nu$.
Similarly, one could use an empirical atmospheric model
such as a linear combination of two of the models above.
In practice the two extra parameters of course give a better fit to the
data, but allow different combinations
of their values and thus do not constrain the fit toward a better
insight into the atmospheric model deficiency.

In principle, one can estimate the value by which the model-predicted
quasi-continuum dry opacity is too high compared to the observations
by analyzing the data taken at very low $pwv$, i.e., where the formal
fit gives negative values for $pwv$. However, this requires a very
precise measurement of the low residual sky-brightness at these low
opacities.  Despite the excellent stability of GREAT, documented by
the Allan-variance minimum times of around 100~sec \citep{hey12},
slight drifts in the total power gain profile caused by, e.g., mechanical
deformation under gravity of the diplexer optics or residual drifts
with temperature in the receiver electronics, limit the precision with
which these measurements can be made.  Figure \ref{fig:load_cal}\
shows that these gain variations on time-scales of several minutes,
i.e., between subsequent hot--cold gain-calibration scans, are on the
order of up to a few percent. These relative errors in the
gain translate into absolute errors in the derived opacities, because the
corresponding sky--hot-observations are affected at the same level.
Note that absolute errors in the opacity of this order are perfectly
tolerable for the calibration of the atmospheric transmission, because they
contribute a small relative error of similar magnitude to the
transmission, and hence to the intensity of the calibrated spectrum.
But they compromise a more precise analysis of the atmospheric model
discrepancies discussed above, because we are talking about small offsets,
i.e., corrections to the $c$-coefficient on the order of a few times
0.01. Only the few sky-measurements immediately following a
load-measurement avoid this problem, but there are too few at very low
opacity within the data set from the SOFIA basic-science flights, and
these are scattered over different receiver tunings, so that they do
not give sufficient statistics to derive a consistent estimate of the
model discrepancy.

\section{Summary and outlook}\label{sec:summary}

Given the problems in the atmospheric models as discussed above, the
GREAT data from the basic science series of flights were calibrated
using individual fits to each receiver band. Although these give
inconsistent values for the best-fit $pwv$ for simultaneous
observations, the relevant entity entering the calibration of the data
is the transmission and not $pwv$. The calibration thus does not take
advantage of the higher signal-to-noise that would be achievable with
a common fit to the full reception bandwidth of the receiver, but it is
nevertheless sufficiently precise to not introduce any significant
errors (depending on the detailed circumstances of the observing
conditions, we estimate the error in the atmospheric transmission thus
derived to be on the order of a few up to a maximum of about 10\%).
Correcting for the atmospheric transmission by applying the
frequency-dependent opacity as given by the atmospheric model, including
the narrow spectral features resulting from the dry atmosphere at full
spectral resolution, is crucial, however, to remove residuals
from these features in the resulting astronomical spectrum, in
particular for absorption measurements against a continuum source.

%
%
%

%
%
%


To improve the calibration, the inconsistencies in the atmospheric
models with regard to the level of broad-band, quasi-continuum opacity
that we identify with the GREAT/SOFIA observations in the THz-regime
have to be overcome by developing a better understanding of the broad-band,
quasi-continuous opacity contributed by the CIA-processes; in
particular, its broad-band variation across the frequency range
covered by the different channels of GREAT needs to be properly
understood. In parallel, a careful analysis of the sky-measurements
available from the basic-science flight series, but also dedicated
future observations tuned to particular frequency settings with
strong, but narrow water lines from the slightly higher atmosphere
that may allow us to independently fix the $pwv$ value and thus to
resolve the ambiguity between dry continuous and wet continuous
opacity contributions, will help to quantify and possibly correct the
discrepancy in the models.

\begin{acknowledgement}

  We thank the SOFIA engineering and operations teams, whose tireless
  support and good-spirit teamwork has been essential for the GREAT
  accomplishments during Early Science, and say Herzlichen Dank to the DSI
  telescope engineering team.

  SOFIA Science Mission Operations are conducted jointly by the
  Universities Space Research Association, Inc., under NASA contract
  NAS2-97001, and the Deutsches SOFIA Institut under DLR contract 50
  OK 0901.

  IRAM (Grenoble, France) has implemented the GREAT/SOFIA specific CLASS header
  extensions, that allow to store the air-borne specific calibration
  parameters within the observation headers.

  Thanks go to Scott Paine, Steve Lord, Nicola Schneider and Juan
  Pardo for help with implementing and using the atmospheric models
  AM, ATRAN, MOLIERE and ATM.

  The development of GREAT was financed by the participating
  institutes, the Max-Planck-Society and the Deutsche
  Forschungsgemeinschaft within the framework of the SFBs 494 and 956.

\end{acknowledgement}





\clearpage \onecolumn
\appendix 

\section{Atmospheric calibration of heterodyne astronomical observations}
\label{app:calib}

(to appear online only)

This appendix is a complete compilation of the calibration scheme for
astronomical observations with a heterodyne receiver, taking into
account the special problems that appear at highest frequencies, i.e.,
at the edge or beyond the validity of the common Rayleigh-Jeans (R-J)
approximation, where the R-J correction across the receiver input band
has to be taken into account with the high IF-bandwidths in use at
present instruments. It also describes a general scheme for
incorporating the correction for atmospheric transmission, which for
observations at high altitudes from the ground, but in particular from
air-borne or balloon-borne altitudes requires a detailed treatment of
the spectrally narrow atmospheric features.

The terminology follows, to the extent possible, the notation defined in
\citet{kut81} and \citet{dow89}.

\subsection{Specific intensity and equivalent brightness temperature}

The specific intensity $I_\nu$ is expressed as an antenna brightness
temperature $T_{A,\nu}$ in the usual radio-astronomical notation,
\begin{equation}
T_{A,\nu}= \frac{\lambda^2}{2  k} I_{\nu}.
\end{equation}
For a black-body radiation field with a physical temperature $T$,
\begin{equation}
B_\nu(T)=
\frac{2h\nu^3}{c^2}\left[
e^{h\nu/kT}-1\right]^{-1},
\end{equation}
the brightness temperature is thus 
\begin{equation}
{\cal J}_\nu(T)=
\frac{\lambda^2}{2{ k}}
B_\nu(T)
=\frac{h\nu}{k}
\left[
e^{h\nu/kT}-1
\right]^{-1}.
\end{equation}
Note that expansion in $\frac{h \nu}{kT} \ll 1$ gives ${\cal
  J}_\nu(T)\approx T-\frac{h\nu}{2k} \left[
  1-\frac{1}{6}\frac{h\nu}{kT} +{\cal O}\left((\frac{h\nu}{kT})^2
  \right)\,\right]\,{}$.  
The difference between two black-bodies at
temperatures $T_2$ and $T_1$ thus is 
\begin{equation}
{\cal J}_\nu(T_2)-{\cal J}_\nu(T_1)
\approx
\left( T_2-T_1 \right)
\left[
1-\frac{1}{6} 
\left( \frac{h\nu}{kT_1}\right)
\left( \frac{h\nu}{kT_2}\right)
\right]=
\left( T_2-T_1 \right)
\left[1-\delta\right]
.
\end{equation} 
With $\frac{h\nu}{k}=48\,{\rm K}\times\nu\,[{\rm THz}]$, the
correction term for a frequency of $\nu=2 {\rm THz}$ and temperatures
of 300~K and 77~K hence is $\delta=-0.066$, small but still
significant, whereas at 345 GHz it would only be $\delta=-0.0019$.

Another quantity of interest is the ratio of black-body intensities
between the image- and signal-sideband. With $\nu_i=\nu_s \pm \nu_{IF}$, 
we obtain 
\begin{equation}
\frac
{{\cal J}_{\nu_i}(T)}
{{\cal J}_{\nu_s}(T)}
\approx 
1 \mp
\left(
1 + \frac{1}{3} \frac{h\nu}{kT}
\right) \frac{h\,\nu_{IF}}{kT},
\end{equation} 
which, for $\nu=2{\rm THz}$, $\nu_{IF}=4 {\rm GHz}$ and a representative 
low temperature of 77~K gives  an upper-/lower-sideband ratio 
of 1.0035, i.e. within 0.35\% of unity. 

\subsection{Antenna temperatures}
We consider the general case of a heterodyne mixer that is sensitive
in both sidebands and looks out to the sky, which emits an intensity
equivalent to an antenna
temperature $T_{A,sky,\nu}$, 
partially picks up intensity from ambient material with
a physical temperature $T_{amb}$, and possibly has a sideband filter that is
terminated at a physical temperature of $T_{term}$. 
The mixer then sees in a
narrow frequency band, centered at a signal sideband frequency $\nu_s$ and an
image sideband frequency $\nu_i$, a total intensity of
\begin{eqnarray}\label{equ:A_tot_int}
T_{A} & = & 
\eta_{mb}\, t_{a,\nu_s} \,
G_s x_s \,T_{MB,\nu_s}
+ \eta_{mb}\, t_{a,\nu_i} \,
G_i x_i\, T_{MB,\nu_i} + \\
&   & (1-f_{amb}) 
\left[ (1-t_{a,\nu_s}) G_s x_s\, 
{\tilde T}_{sky,\nu_s}+ 
        (1-t_{a,\nu_i}) G_i x_i 
\,{\tilde T}_{sky,\nu_i}
\right] +
        \nonumber  \\
&   & f_{amb}
 \left[ G_s x_s\, {\cal J}_{\nu_s}(T_{amb})
+G_i x_i\,{\cal J}_{\nu_i}(T_{amb}) \right]
+ \nonumber \\&&
(1-x_s) G_s \,{\cal J}_{\nu_s}(T_{term}) 
+
(1-x_i) G_i \, {\cal J}_{\nu_i}(T_{term}), \nonumber  
\end{eqnarray}
where we have introduced the following definitions:
\begin{eqnarray}
\label{equ:app-Ta}
f_{amb} && {\rm fraction\ of\ ambient\ material\ in\ sky\ beam} \\
\eta_{mb} && {\rm main\ beam\ efficiency}\nonumber \\
T_{mb,s,i} && {\rm source\ main\ beam\ brightness\ temperature\ 
in\ signal(image)\ sideband}\nonumber \\
{\tilde T}_{sky,\nu} && {\rm effective\ sky\ brightness\ 
temperature\ at\ frequency\ } \nu,\nonumber \\
 &&  {\tilde T}_{sky,\nu}:=T_{A,sky,\nu}/(1-t_{a,\nu})\nonumber \\
t_{a,\nu} && {\rm atmospheric\ transmission\ at\ frequency\ }\nu \nonumber \\
G_{s,i} && {\rm signal(image)\ sideband\ gain\!\!:\ }\,G_s+G_i=1\nonumber \\
x_{s,i} && {\rm signal(image)\ response\ in\ sky\ beam,}\nonumber \\
 && {\rm with\ coupling\ to\ sideband\ filter\ termination\!\!:\ } (1-x_{s,i})\nonumber.
\end{eqnarray}
Here, we have assumed that the coupling factor to ambient and the main
beam efficiency are independent of the reception sideband.

\subsection{Types of receivers}
We can distinguish the following two cases straightforwardly:

\begin{itemize}
\item
case A: a receiver with no sideband filter has
\begin{equation} 
          \begin{array}{l}
             x_s=x_i=1 \nonumber \\ \label{caseA}
             \Rightarrow G_sx_s+G_ix_i
=G_s+G_i=1 \,\,
\Rightarrow 
G_s=1-G_i=\frac{1}{1+G_i/G_s}. \nonumber\\
          \end{array}
\end{equation}
GREAT on SOFIA falls into this category.

\item
case B: a receiver with a sideband filter, i.e., non-zero coupling to the
sideband termination, has
\begin{equation} \nonumber
      x_s+x_i=1\,\,
 \Rightarrow x_s=1-x_i=\frac{1}{1+x_i/x_s}.
   \end{equation}
\end{itemize}
Other hardware implementations correspond to different values of $G_{s,i}$ and $x_{s,i}$.

\subsection{Main-beam brightness temperature and forward beam brightness temperature}
In the above, we have expressed the source signal as a {\it main-beam
  brightness temperature}, i.e., as the convolution of the source
brightness distribution on the sky with the main-beam profile of the
telescope. The coupling of the detector to the telescope main beam is
the main-beam efficiency $\eta_{mb}$. This is appropriate if the
telescope has a clean main beam, thus avoiding additional pick-up of
source intensity in side-lobes (error-beam pickup). This is also a
good description if the source is compact so that it has only
intensity within the main beam.

Alternatively, the source signal can be expressed as a forward beam
brightness temperature $T^{*}_{A,\nu}$, i.e., as the convolution of
the source brightness with the full forward antenna response. The
relevant coupling efficiency is then the forward efficiency
$\eta_{fwd}$, which in the above nomenclature is
$\eta_{fwd}=1-f_{amb}$. The term $\eta_{mb}\,T_{MB,\nu}$ then is
replaced by $\eta_{fwd}\,T^{*}_{A,\nu} = (1-f_{amb})\,T^{*}_{A,\nu}$.

\subsection{Sky brightness and opacity}
\subsubsection{Opacity}
The sky transmission along a line-of-sight at elevation $El$ for a
plane-parallel atmosphere, i.e.,
approximating the airmass by $A=1/\sin(El)$ (the latter assumption
being valid except at extremely low elevations), is
\begin{equation}
t_{a,\nu}=e^{-\tau_\nu/\sin(El)}
\end{equation}
is given by the zenith opacity $\tau_\nu$, where we drop the index $Z$
in the following.

The zenith-opacity of a particular species $\xi$ is the integral along
the line-of-sight from the observers altitude $h$ to space over the
abundance of that species times its absorption coefficient. The
abundance at altitude $s$ is expressed as the volume-mixing ratio
\begin{equation}
x_\xi(s) =
{n_\xi(s)}/{n(s)}
\end{equation} 
relative to the total density $n(s)$ (we closely follow the
nomenclature of the AM handbook \citep{am}). The general expression
for the opacity of a particular atmospheric species $\xi$ with number
density $n_\xi(s)=x_\xi(s)\,n(s)$ is then the line-of-sight integral
over the product of its abundance times its pressure-, $P(s)$ , and
temperature-, $T(s)$, -dependent molecular absorption coefficient,
\begin{equation}
\tau_{\nu,\xi}(h)=
\int_h^\infty x_\xi(s)\, n(s)
\,k_{\xi}(\nu,P(s),T(s))\,ds.
\end{equation} 
The total opacity is obtained by summation over all species, 
\begin{equation}
\tau_\nu(h)=\sum_\xi \tau_{\nu,\xi}(h).
\end{equation} 
The details of the microphysics, i.e., the kind of absorption (single
line transitions, collision-induced absorption, etc.) are contained in
the molecular absorption coefficients. For collision-induced
absorption (CIA) the coefficient is proportional to the the abundance
of the collision partner(s) $\hat \xi$, i.e.,
\begin{equation}
k_\xi=x_{\hat \xi}(s)\,n(s)\,
\kappa^{CIA}_{\xi,{\hat \xi}}(\nu,P(s),T(s)),
\end{equation} 
which for self-broadening ($\xi=\hat \xi$) leads to a quadratic
dependence on the abundance of the species.

Instead of altitude, the equivalent pressure can be used to specify the height in the atmosphere (pressure altitude) through the hydrostatic equation 
\begin{equation}
n(h)\,dh
=
\frac{dP}{m_{air}\,g},
\end{equation} 
with the mean molecular weight $m_{air}$ of the atmosphere and the
gravitational acceleration $g$.

We will show below that the zenith opacity of
the atmosphere above the observer can be separated into a wet and a
dry part (where the dry part may contain opacity from the high water
column in the stratosphere with constant volume mixing ratio that
does not change with local humidity and daily weather). Thus, the
opacity can be written as
\begin{equation}\label{equ:A_bc}
\tau_\nu(h)=b_\nu(h)\,pwv + c_\nu(h).
\end{equation}

\subsubsection{Sky brightness}

The sky brightness is the integral along the line-of-sight at
elevation $El$ over the source function times the attenuation,
which for a plane-parallel layering, approximating the airmass as
$A=1/\sin(El)$, reads
\begin{equation}\label{equ:A_sky}
T_{A,sky,\nu}
=
\int_0^{\tau_\nu}{\cal J}\left(T(\tau_\nu ')\right)\,
\exp{\left(-\tau_\nu '/\sin El\right)}\,d\tau_\nu '/\sin El.
\end{equation}
From this, the {\it effective sky brightness} as defined in 
equ.~\ref{equ:app-Ta}\  can be calculated with a given line-of-sight
transmission $t_{a,\nu}$ as
\begin{equation}\label{equ:A_sky_eff}
{\tilde T}_{sky,\nu}=T_{A,sky,\nu}/\left( 1- t_{a,\nu} \right).
\end{equation}

\subsection{Parameterization of the atmospheric water vapor}
For water, a reasonable altitude distribution in the stratosphere and upper
troposphere is given by a low, constant mixing ratio, dominant in the upper
stratosphere, plus an exponential term, extending through the upper
troposphere and lower stratosphere \citep{mas68,her09}. We thus write
\begin{eqnarray}\label{equ:A_h2o_col}
x_{{\rm H}_2{\rm O}}(s)&=&x_c+
x_e\,\exp\left[ -(s-h_e)/h_s \right]\\
&=&
x_c + {x}_e' \exp(-s/h_s), \nonumber
\end{eqnarray}
with $x_e'=x_e\exp(h_e/h_s)$. The typical value for $x_c$ is
$2.5\,10^{-6}$ \citep{mas68,her09}; $h_c$ or $x_e'$ can be
adjusted to match the desired column of water vapor above the
observer. Equ.\ \ref{equ:A_h2o_col}\ gives a total column density of water above
altitude $h$ of
\begin{eqnarray}
N_{{\rm H}_2{\rm O}}(h)&=&
\int_h^\infty \left( x_c+x_e\,
\exp \left[ -(s-h_e)/h_s\right]
\right)\,
n(s)\,ds\\
&=&
x_c \int_h^\infty n(s)\,ds +
x_e \exp(h_e/h_s) \int_h^\infty
\exp(-s/h_s)\,n(s)\,ds\nonumber \\
&=&
x_c\,N_c(h)+x_e' \,N_e(h),\nonumber
\end{eqnarray}
using the abbreviations 
\begin{eqnarray}\label{equ:A_abbrev}
N_c(h)&=&\int_h^\infty n(s)\,ds \\
N_e(h)&=&\int_h^\infty e^{-s/h_s}\,n(s)\,
ds. \nonumber
\end{eqnarray}
The minimum column is given by the constant volume mixing ratio, setting
\begin{equation}
N_{{\rm H}_2{\rm O},{min}}=x_c\,N_c(h).
\end{equation} 
Hence, 
\begin{equation}
x_e'=
\left(N_{{\rm H}_2{\rm O}}(h)
-x_c\,N_c(h)\right)/N_e(h).
\end{equation} 
Similarly, using the abbreviations 
\begin{eqnarray}
K_c(h)&=&\int_h^\infty n(s)\,
k_{{\rm H}_2{\rm O}}(\nu,P(s),T(s))\,ds\\
K_e(h)&=&\int_h^\infty n(s)\,
e^{-s/h_s}\,
k_{{\rm H}_2{\rm O}}(\nu,P(s),T(s))\,ds\nonumber \\
K_{ee}(h)&=&\int_h^\infty n^2(s)\,
e^{-2s/h_s},
\kappa^{self}_{{\rm H}_2{\rm O}}
(\nu,P(s),T(s))\,ds\nonumber \\
K_{cc}(h)&=&\int_h^\infty n^2(s)\,
\kappa^{self}_{{\rm H}_2{\rm O}}
(\nu,P(s),T(s))\,ds\nonumber \\
K_{ec}(h)&=&\int_h^\infty n^2(s)\,
e^{s/h_s},
\kappa^{self}_{{\rm H}_2{\rm O}}
(\nu,P(s),T(s))\,ds,\nonumber 
\end{eqnarray}
the water opacity is 
\begin{eqnarray}
\tau_{\nu,{\rm H}_2{\rm O}}(h)
&=&
x_c\,\left\{K_c(h)
+x_c\,K_{cc}(h)\right\}
\\&&+
x_e'
\left\{K_e(h)+x_e'\,K_{ee}(h)
+x_c \,K_{ec}(h)\right\}\nonumber \\
&\approx&
x_c\,K_c(h)+x_e'\,K_e(h). \nonumber
\end{eqnarray}
Note that in this notation, the absorption coefficient of water includes the
contributions from CIA-processes due to collisions with other species,
\begin{equation}
k_{{\rm H}_2{\rm O}}(\nu,P(s),T(s)) \rightarrow
k_{{\rm H}_2{\rm O}}(\nu,P(s),T(s))
+ \sum_{\xi \ne {\rm H}_2{\rm O}}
x_{\xi}\,n(s)\,\kappa_{{\rm H}_2{\rm O},\xi}(\nu,P(S),T(s)).
\end{equation}

The self-broadening contribution for water in the thin, upper
atmosphere can be neglected
\citep{bau87,bau89}. Combining the
opacity contribution from the constant volume-mixing ratio part of
water, $\tau_{c,\nu}(h)=x_c\,K_c(h)$, with the opacity of all other
species into
\begin{equation}
\label{equ:c_h}
c_\nu(h)=\tau_{c,\nu}(h)+\sum_{\xi\ne{\rm H}_2{\rm O}} \tau_{\nu,\xi}(h),
\end{equation} 
we can write the zenith opacity as separated into a {\it dry opacity},
resulting from all absorption from species other than water, but
including the constant mixing ratio contribution from water in the
upper stratosphere, and a {\it wet opacity}, which increases in proportion
with the precipitable water vapor from the exponential part of the
water vapor above the observer,
\begin{equation}
pwv=
\frac{\rho_{{\rm H}_2{\rm O}}}
{m_{{\rm H}_2{\rm O}}}\,
\left(
N_{{\rm H}_2{\rm O}}(h)-
N_{{\rm H}_2{\rm O},min }
\right).
\end{equation} 
Hence, we obtain equ.\ \ref{equ:A_bc} with
\begin{equation}
\label{equ:b_h}
b_\nu(h)=
\frac{
m_{{\rm H}_2{\rm O}}
}{
\rho_{{\rm H}_2{\rm O}}
}
\frac{1}{N_e(h)}
\,K_e(h).\nonumber
\end{equation} 

\subsubsection{Alternative parameterization of the atmospheric water vapor}
Equ.~\ref{equ:A_h2o_col} is a rather ad hoc parameterization,
which approximates the water abundance altitude profile above the
SOFIA flight altitudes.
An alternative parameterization, which is actually the one
used in the present implementation with the {\it kalibrate} task in the
{\it kosma\_software} package, would be as
follows:
\begin{eqnarray}\label{equ:A_h2o_alt}
x_{{\rm H}_2{\rm O}}(s)&=&
\left\{
\begin{array}{ll}
x_c, & h>h_e\\
x_c\,e^{-(s-h_e)/h_s)}, & h\le h_e
\end{array}
\right. \\
&=&
x_c\,e^{-(s-h_e)/h_s)}
+x_c\,
\left\{
\begin{array}{ll}
1-e^{-(s-h_e)/h_s)}, 
& h>h_e\nonumber \\
0, & h\le h_e
\end{array}
\right. \nonumber
\end{eqnarray}
This gives, 
using
$N_e(h)$ and $N_c(h)$ as defined in equ.\ \ref{equ:A_abbrev} above
\begin{eqnarray}
N_{{\rm H}_2{\rm O}}(h)&=&
x_c\,e^{h_e/h_s}
\left\{
\int_h^\infty e^{-\frac{s}{h_s}}\,n(s)\,ds+
e^{-\frac{h_e}{h_s}} 
\int_{h_e}^\infty
\left(1-e^{-\frac{s-h_e}{h_s}}\right)
\,n(s)\,ds
\right\}\\
&=&
{x}_c\,e^{h_e/h_s}
\left\{
N_e(h)-N_e(h_e)+e^{-h_e/h_s}N_c(h_e)
\right \}. \nonumber
\end{eqnarray}
As in the case above, $h_e$ can be varied to scale the water profile
to the desired total column of water above the observer. Owing to the
different parameterization, the variation of $N_{{\rm H}_2{\rm O}}(h)$
with $h_e$ is now a more complex dependence and solving for $h_e$ is
not possible analytically. The minimum column of water above the
observer is now given by the contribution with the constant volume
mixing ratio when $h_e=h$, namely
\begin{equation}
\left.
N_{{\rm H}_2{\rm O}}\right|_{h_e=h}
=x_c \int_h^\infty n(s)\,ds
=x_c\,N_c(h),
\end{equation} 
identical to the case above. Similarly, and again neglecting the
self-broadening contributions of water, the water opacity is now given
by
\begin{equation}\label{equ:A_h2o_tau1}
\tau_{\nu,{\rm H}_2{\rm O}}=
x_c\,e^{h_e/h_s}
\left\{
K_e(h)-K_e(h_e)+e^{-h_e/h_s}K_c(h)
\right\}.
\end{equation} 
For small increases above the minimum, i.e.\ $h_e=h+\epsilon$, we have 
\begin{eqnarray}
\label{equ:A_h2o_tau3}
\left.
N_{{\rm H}_2{\rm O}}\right|^
{h_e=h+\epsilon}_
{h_e=h}
&=&
x_c  e^{(h+\epsilon)/h_s}
\left(
N_e(h)-N_e(h+\epsilon) \right)\\
&&
+x_c\left(N_c(h_e)
-N_c(h) \right)
\nonumber \\
&=&
x_c\,e^{(h+\epsilon)/h_s}\,
\int_h^{h+\epsilon}n(s)\,ds
-x_c \int_h^{h+\epsilon}n(s)\,ds\nonumber \\
&\stackrel {\epsilon \to 0}{\rightarrow} &
x_c\,e^{h/h_s}\,
\left(1+\frac{\epsilon}{h_s} \right) \,
n(h)\,\epsilon-x_c\,n(h)\,\epsilon\nonumber \\
&=&
x_c\frac{n(h)}{h_s}\,\epsilon^2. \nonumber
\end{eqnarray}
For the water opacity we obtain, following equ.\ \ref{equ:A_h2o_alt}, 
and abbreviating
$k(s)= k_{{\rm H}_2{\rm O}}(\nu,P(s),T(s))$,
\begin{eqnarray}\label{equ:A_h2o_tau2}
\left.
\tau_{\nu,{\rm H}_2{\rm O}}\right|^
{h_e=h+\epsilon}_
{h_e=h}
&=&
x_c  e^{(h+\epsilon)/h_s}
\left(
K_e(h)-K_e(h+\epsilon) \right)\\
&&
+K_c\left(N_c(h_e)
-K_c(h) \right)
\nonumber \\
&=&
x_c\,e^{(h+\epsilon)/h_s}\,
\int_h^{h+\epsilon}n(s)\,k(s)\,ds
-x_c \int_h^{h+\epsilon}n(s)\,k(s)\,ds\nonumber \\
&\stackrel {\epsilon \to 0}{\rightarrow} &
x_c\,e^{h/h_s}\,
\left(1+\frac{\epsilon}{h_s} \right) \,
n(h)\,k(h)\,\epsilon-x_c\,n(h)\,k(h)\,\epsilon\nonumber \\
&=&
x_c\frac{n(h)}{h_s}\,k(h)\,\epsilon^2, \nonumber
\end{eqnarray}
so that the opacity is proportional to $pwv$, with
\begin{equation}\label{equ:A_prop}
b_\nu (h)=
\frac{
\rho_{{\rm H}_2{\rm O}}
}{
m_{{\rm H}_2{\rm O}}
}
\,
k_{{\rm H}_2{\rm O}}(\nu,P(h),T(h)).
\end{equation}
For larger water columns, the opacity is not necessarily proportional
to $pwv$: although the $N_e$-, resp.\ $K_c$-, contribution in equs.\
\ref{equ:A_h2o_tau3}\ and \ref{equ:A_h2o_tau2}\ becomes rapidly insignificant relative to the first terms
because of the exponential decrease with increasing $h_e$: the variation
of the altitude distribution due to the changing point of intersection
between the constant and exponential part, $h_e$, results in changing
line profiles in the first terms. However, as long as the variation of
the line profile with altitude is relatively slow, i.e.\
$k_\nu(s)\approx k_\nu(h)$, we approximately have 
\begin{equation}
K_e(h)-K_e(h_e)\approx k_\nu(h)\left( N_e(h)-N_e(h_e)\right),
\end{equation} 
so that the
proportionality between opacity and water column holds again, with
equ.\ \ref{equ:A_prop} holding approximately.

\subsection{Spectrometer response and gain profile}
The spectrometer detects in each frequency channel (we omit the index
specifying the spectrometer channel, resp.\ intermediate frequency, $\nu_{IF}$,
in the following) a count rate (counts/sec) $C$ proportional to the
radiation input plus the receiver noise (expressed as an equivalent
brightness temperature at the receiver input, $T_{rec}$), the total being
\begin{equation}\label{equ:A_tsys}
T_{sys}=T_{rec}+T_A.
\end{equation} 
Additionally, depending on the detailed detection mechanism of the
power in the spectrometer channels, the receiver may have a
count-offset $Z$, i.e., a non-zero count rate at zero (IF-) input
power.  The detected count rate is thus
\begin{equation}\label{equ:A_counts}
C=\gamma (T_{A}+T_{rec})+Z,
\end{equation}
so that the spectrometer gain $\gamma$ (gain profile, if specified as
a function of frequency) determines the response of the receiver for
any difference measurement:
\begin{equation}
\Delta C =C_2-C_1=\gamma(T_{A,2}-T_{A,1})=
\gamma \,\Delta T_A.
\end{equation} 

\subsection{Gain calibration: Hot- and cold-load measurements}
For calibration measurements on the hot- or cold-load, the mixer sees 
\begin{eqnarray}\label{equ:A_load}
T_{A,{}^{hot}_{cold}}
&=&
(G_sx_s+G_ix_i)
T'_{{}^{hot}_{cold}}
+\\&&
(1-x_s)
G_s
{\cal J}_{\nu_s}(T_{term})
+(1-x_i)
G_i
{\cal J}_{\nu_s}(T_{term}), \nonumber
\end{eqnarray}
where we have used the abbreviation
\begin{equation}
T'_{{}^{hot}_{cold}}=\frac{
G_sx_s
{\cal J}_{\nu_s}(T_{{}^{hot}_{cold}})
+
G_ix_i
{\cal J}_{\nu_i}(T_{{}^{hot}_{cold}})
}
{
G_sx_s(1+\Gamma)
}.
\end{equation}
With the known physical temperatures of the loads, and the known
sideband gains and sideband filter coupling coefficients, the
spectrometer gain  in each spectral resolution element  
can be measured from the difference in count rates
between the hot- and cold-load measurement:
\begin{equation}
\gamma=\frac{C_{hot}-C_{cold}}
{T_{A,hot}-T_{a,cold}}
=
\frac{C_{hot}-c_{cold}}
{(G_s x_s +G_ix_i)
\left((T'_{hot}-T'_{cold}\right)
}.
\end{equation} 
As discussed above, the difference between signal- and image-sideband
brightness temperatures may be neglected to relative high precision,
resulting in $T'_{{}^{hot}_{cold}}={\cal J}_\nu(T_{{}^{hot}_{cold}})$,
where an obviously consistent choice for the frequency at which to
evaluate the brightnesses would be
$\nu=\frac{1}{2}(\nu_s+\nu_i)=\nu_{LO}$.

\subsection{Sky transmission calibration}
The sky transmission is derived from measurements on the blank sky
(either explicit sky-measurements, or the off-source positions of an
astronomical observation). The measured difference between sky- and
hot-load measurement gives
\begin{equation}
\left.\left(T_{A,sky}-T_{A,hot}\right)\right|_
{meas}
=\frac{C_{sky}-C_{hot}}
\gamma
\end{equation} 
(note that here and in the following, the cold-load measurement and
temperatures may be used instead of the hot-load).  This is compared
with the theoretical expression for the brightness difference, which,
using equations \ref{equ:A_tot_int}\ and \ref{equ:A_load}, gives
\begin{eqnarray} 
\left.
\left(T_{A.sky}-T_{A,hot}\right)\right|_{meas} & = & 
(1-f_{amb})\times\\&& 
\left[ (1-t_{a,\nu_s}) G_s x_s\, 
{\tilde T}_{sky,\nu_s}+ 
        (1-t_{a,\nu_i}) G_i x_i 
\,{\tilde T}_{sky,\nu_i}
\right] +
         \nonumber \\
&   & f_{amb}
 \left[ G_s x_s\, {\cal J}_{\nu_s}(T_{amb})
+G_i x_i\,{\cal J}_{\nu_i}(T_{amb}) \right]
- \nonumber \\&&
\left[
G_sx_s{\cal J}_{\nu_s}(T_{hot})
+G_ix_i{\cal J}_{\nu_i}(T_{hot})\right]\nonumber \\
&=&
(1-f_{amb})
\left\{
G_s x_s
\left[{\tilde T}_{sky,\nu_s}(1-t_{a,\nu_s})\right.
-{\tilde T}_{hot,\nu_s}\right]
\nonumber \\ && \left.\hspace{1.5cm}+
G_i x_i
\left[{\tilde T}_{sky,\nu_i}(1-t_{a,\nu_i})
-{\tilde T}_{hot,\nu_i}\right]
\right\}, \nonumber
\end{eqnarray}
where we have used the abbreviation for the {\it effective hot-temperature}
\begin{equation}
{\tilde T}_{hot,\nu}=
\frac{{\cal J}_\nu(T_{hot})-
f_{amb}\,{\cal J}_\nu(T_{amb})}
{1-f_{amb}}.
\end{equation} 
Introducing the additional approximations
\begin{equation}
\label{equ:abbrev}
a_{s}=
\frac{{\tilde T}_{sky,\nu_{s}}}
{{\tilde T}_{hot,\nu_{s}}},\,
a_{i}=
\frac{{\tilde T}_{sky,\nu_{i}}}
{{\tilde T}_{hot,\nu_{i}}},
\end{equation}
\begin{eqnarray*}
\Gamma&=&\frac{G_ix_i}{G_sx_s}
\end{eqnarray*}
\begin{eqnarray*}
\omega&=&
\frac{{\tilde T}_{sky,\nu_{i}}}
{{\tilde T}_{sky,\nu_{s}}}\\
\end{eqnarray*}
\begin{eqnarray*}
\Omega&=&\frac
{{\tilde T}_{hot,\nu_{i}}-
{\tilde T}_{sky,\nu_{i}}}
{{\tilde T}_{hot,\nu_{s}}-
{\tilde T}_{sky,\nu_{s}}}
\end{eqnarray*}
gives 
\begin{eqnarray}
\label{equ:A_skybrightness}
\left.\left(
T_{A,sky}-T_{A,hot}\right)\right|_{meas} & = & 
-(1-f_{amb})\,
{\tilde T}_{hot,\nu_s}\,a_s\,
G_sx_s\,(1+\Gamma)\times  \\
&&
\left\{
t_{a,\nu_s}
\underbrace{
\frac{1+\Gamma\,\omega\,
\left( t_{a,\nu_i}/t_{a,\nu_s}\right)}
{1+\Gamma}}_{1/\alpha}
+
\frac{1-a_s}{a_s}\,
\frac{1+\Gamma\,\Omega}{1+\Gamma}
\right\}\nonumber \\
 & = & 
-(1-f_{amb})\,
{\tilde T}_{sky,\nu_s}\
G_sx_s\times\nonumber \\
&&
\left\{
t_{a,\nu_s}
\left( 1+\Gamma\,\omega\,
\left( t_{a,\nu_i}/t_{a,\nu_s}\right)
\right)
+
\frac{1-a_s}{a_s}\,
\left(1+\Gamma\,\Omega\right)
\right\}.  \nonumber
\end{eqnarray}
As discussed above, $\omega \approx 1$ to high precision
for the relevant range of frequencies and temperatures. $\Omega
\approx 1$ only holds as long as the effective sky brightness is
close to a black-body function, i.e., ${\tilde T}_{sky,\nu}\approx
{\cal J}_\nu(T_{sky})$ , with a frequency-independent effective
physical sky temperature $T_{sky}$. This is not necessarily the case:
a narrow-line absorption in the signal band resulting from water at
the low temperatures of the upper troposphere and a similarly narrow
feature from a high-altitude dry species at the higher physical
temperature of the upper stratosphere may have very different
effective sky temperatures, following their definitions according to
equations \ref{equ:A_sky}\ and \ref{equ:A_sky_eff}.

Assuming for simplicity $\Omega=1$ and $\omega=1$, and considering a
double-sideband receiver with equal gain in both sidebands and no
sideband filter, i.e., $\Gamma=1$, $x_s=1$ and $G_s=1$, we obtain the
simple formula
\begin{eqnarray} 
\left.\left(
T_{A,sky}-T_{A,hot} 
\right)\right|_{meas }& = & 
-(1-f_{amb})\,
{\cal J}_{\nu_s}(T_{sky})\,
\left\{
\frac{1}{2}
\left(
t_{a,\nu_s} + t_{a,\nu_i}
\right)
+ \frac{1-a_s}{a_s}\right\}.
\end{eqnarray}
For a single-sideband receiver, i.e., $G_i=0$, and hence
$\Gamma=0$,$x_s=1$ and $G_s=1$, we obtain
\begin{eqnarray} 
\left.\left(
T_{A,sky}-T_{A,hot} 
\right) \right|_{meas}& = & 
-(1-f_{amb})\,
{\cal J}_{\nu_s}(T_{sky})\,
\left\{
t_{a,\nu_s} 
+ \frac{1-a_s}{a_s}\right\}.
\end{eqnarray}
The latter can be immediately inverted to calculate
the sky transmission in the signal band; in the former case, only the
average transmission in the signal- and image-sideband can be
determined.  

The same holds for the general case of equation \ref{equ:A_skybrightness},
which gives
\begin{equation}\label{equ:A_model_trans}
t_{a,\nu_s} +\Gamma\,\omega\,
t_{a,\nu_i}
=
\frac{
(C_{hot}-C_{sky})/\gamma}
{(1-f_{amb})\,{\tilde T}_{sky,\nu_s}\,
G_sx_s}
-
\frac{1-a_s}{a_s}\,
\left(1+\Gamma\,\Omega\right),
\end{equation} 
from which the sky transmission in the signal-sideband alone can be
determined only with additional knowledge on i) the ratio of signal-
and image-sideband transmission, and ii) the effective sky brightness
at the signal- and image-band frequency, ${\tilde T}_{sky,\nu_{s,i}}$,
which is also needed to calculated $\Omega$ with full precision.

Determining the transmission across
the spectrometer channel by channel is not the best strategy, because
the values derived would be very noisy due to the relatively low
signal-to-noise ratio per spectrometer channel. This can be improved
by using broader frequency averages, but at the cost of loosing the
information of variations of the atmospheric absorption with frequency
on frequency scales below the averaging interval.  

To overcome the difficulty of the unknown atmospheric transmission
ratio between image- and signal-sideband, one has to use additional
knowledge about the atmosphere by applying an atmospheric model and
fitting the observed sky-hot brightness to the model-predicted one.

This is implemented in the KOSMA-observing software package {\it
  kosma\_software} in the task {\it kalibrate} as follows: (i) the
first step is to identify and discard ``bad'' channels, e.g., at the
band edges without significant IF-response and saturated channels
etc. (ii) the observed and calibrated sky-hot difference across the
reception bandwidth of the receiver is then fitted with the
atmospheric model prediction, following equ.\ \ref{equ:A_skybrightness}, 
using the $pwv$ as a
free parameter; the implementation at present makes the simplifying
assumption of $\omega=1$ and $\Omega=1$.  The best-fit value is thus
determined with high precision, making use of the full spectral
information observed, i.e., with a good signal-to-noise ratio.
The user can control whether the fit is performed as a ``common'' fit to all
simultaneously observed spectra (several receiver pixels at several
frequency bands) or individually for each spectrometer.  (iii) The
measured on-off source spectrum can then be calibrated on the
main-beam brightness temperature scale via
\begin{equation}
\Delta T_{mb,SSB} = 
\frac{C_{ON}-C_{OFF}}
{\eta_{mb}\,\gamma\,
G_sx_s}
\exp{\left(b_\nu\,pwv+c_\nu\right)}.
\end{equation} 
This is made separately for each spectrometer channel, thus
applying the frequency dependence of the atmospheric transmission as
specified by the atmospheric model to the data.

\subsection{Determination of receiver noise and system temperature }
Equs.~\ref{equ:A_tsys}\ and \ref{equ:A_counts} define the system
temperature used for proper noise-weighting of the data when
averaging, and the receiver temperature that needs to be optimized when
tuning the receiver, both on a $T_A$ scale, however. Comparison with
equ.~\ref{equ:A_load}\ shows that the system temperature and receiver
temperature referred to an equivalent black-body with physical
temperature $T_{sys,phys}$, resp. $T_{rec,phys}$ at the receiver
input, either double-sideband or single-sideband, is
\begin{equation}
  T_{rec,phys,DSB}=\frac{T_{rec}}{G_sx_s+G_ix_i} \qquad \qquad
  T_{rec,phys,SSB}=\frac{T_{rec}}{G_sx_s}
\end{equation}
respectively 
\begin{equation}\label{equ:A_tsys_dsb}
  T_{sys,phys,DSB}= \frac{T_{sys}}{G_sx_s+G_ix_i} \qquad \qquad
  T_{sys,phys,SSB}= \frac{T_{sys}}{G_sx_s}.
\end{equation}
From the defining equ.~\ref{equ:A_counts}, 
we obtain for the hot- and cold-load measurement 
\begin{eqnarray}
C_{hot}-z & = & \gamma\,(T_{rec}+T_{A,hot})\\
C_{cold}-z & = & \gamma\,(T_{rec}+T_{A,cold}), \nonumber
\end{eqnarray}
and can derive (Y-factor method) 
\begin{equation}
  Y=\frac{C_{hot}-z}{C_{cold}-z}=\frac{T_{rec}+T_{A,hot}}{T_{rec}+T_{A,cold}}.
\end{equation}
Solving for $T_{rec}$ gives 
\begin{eqnarray}
  T_{rec} & = & \frac
{T_{A,hot}-YT_{A,cold}}{Y-1}\\
                      & = & \underbrace
{\frac{
T'_{hot}-
Y\,
T'_{cold}
}{Y-1}}_{T'_{rec}}
\left(G_sx_s+G_ix_i\right)
\nonumber \\
&&-\left[
G_s(1-x_s)
{\cal J}_{\nu_s}(T_{term})
+
G_i(1-x_i)
{\cal J}_{\nu_i}(T_{term})
\right]\nonumber \\
                      & = & T'_{rec}
(G_sx_s+G_ix_i)
-\left[
G_s(1-x_s)
{\cal J}_{\nu_s}(T_{term})
+
G_i(1-x_i)
{\cal J}_{\nu_i}(T_{term})
\right], \nonumber
\end{eqnarray} 
so that 
\begin{equation}
    \begin{array}{rcl}
        T_{rec,phys,DSB} & = & 
T'_{rec}-
\frac{
G_s(1-x_s){\cal J}_{\nu_s}(T_{term})
+G_i(1-x_i){\cal J}_{\nu_i}(T_{term})}
{G_sx_s+G_ix_i}
\\
\\
T_{rec,phys,SSB} & = & T'_{rec}
(1+\Gamma)-
\frac{
G_s(1-x_s){\cal J}_{\nu_s}(T_{term})
+G_i(1-x_i){\cal J}_{\nu_i}(T_{term})}
{G_sx_s}
\end{array}.
\end{equation}
Ignoring the difference in brightness between the signal- and
image-sideband on the hot- and cold load, $T'_{rec}$ reduces to the standard
formula for the receiver temperature in the y-factor method,
\begin{equation}
T'_{rec}\approx 
\frac{{\cal J}_\nu(T_{hot})-
Y\,{\cal J}_\nu(T_{cold})}
{Y-1}.
\end{equation} 
The additional term on the right-hand-side corrects for the coupling
to the image-sideband termination in the sideband filter and reduces
to $0$ with $x_s=x_i=1$.  

We obtain for the system temperature (on the
off-source position on the sky)
\begin{eqnarray}
  T_{sys} & = & T_{rec}+T_{A,sky}\\
& = &
\left(T_{rec} +T_{A,hot}\right)+
\left(T_{A,sky}-T_{A,hot}\right) \nonumber \\
&=&
\left(T'_{rec}+T'_{hot}\right)
G_sx_s(1+\Gamma)
+\left(T_{A,sky}-T_{A,hot}\right). \nonumber 
\end{eqnarray}
Insertingly, we obtain from equ.~\ref{equ:A_skybrightness}
\begin{eqnarray}
T_{sys}&=&G_sx_s(1+\Gamma) \times
\\&&
\left[
T'_{rec}+T'_{hot}-
{\tilde T}_{sky,\nu_s}\,(1-f_{amb})
\left(
t_{a,\nu_s}\frac{1}{\alpha}
+\frac{1-a_s}{a_s}\,\frac{1+\Gamma\,
\Omega}{1+\Gamma}
\right)
\right], \nonumber
\end{eqnarray}
and $T_{sys,DSB}$ or $T_{sys,SSB}$ following equ.~\ref{equ:A_tsys_dsb}.


\begin{thebibliography}{}
\bibitem[Bauer et al.(1987)]{bau87} Bauer, A., Godon, M., Kheddar, M. et al.
    1987, JQSRT, 37, 531
\bibitem[Bauer et al.(1989)]{bau89} Bauer, A., Godon, M., Kheddar, M. et al.
    1989, JQSRT, 41, 49
\bibitem[Becklin \& Gehrz(2009)]{bec09} Becklin, E. E. \& Gehrz, R. D.
    2009, ASPC, 417, 101
\bibitem[Camy-Peyret et al.(2004)]{cam04} Camy-Peyret, C., Payan, S.,
    Dufour, G. et al. 2004, ESASP, 562, 31
\bibitem[Downes(1989)]{dow89} Downes, D. 1989, LNP, 333, 351
\bibitem[Emmons et al.(1994)]{emm94} Emmons, L. K., Reeves, J. M.,
    Shindell, D. T. et al. 1994, OZTS, p. 543
\bibitem[G\"usten et al.(2006)]{gues06} G\"usten, R., Nyman, L. \AA.,
    Schilke, P. et al. 2006, \aap, 454, 13
\bibitem[Herbin et al.(2009)]{her09} Herbin, H., Hurtmans, D.
    \& Clerbaux, C. 2009, ACP, 9, 9433
\bibitem[Heyminck et al.(2012)]{hey12} Heyminck, S., Graf, U. U.,
    G\"usten, R. et al. 2012, \apj
\bibitem[Kawamura et al.(2005)]{kaw05} Kawamura, A., Mizuno, N.,
    Yonekura, Y. et al. 2005, IAUS, 235, P275
\bibitem[Kutner \& Ulich(1981)]{kut81} Kutner, M. L. \& Ulich, B. L.
    1981, \apj, 250, 341
\bibitem[Pardo et al.(2001)]{atm} Pardo, J. R., Cernicharo, J.
    \& Serabyn, E. 2001, IEEE Trans. on Antennas and Propagation, 49/12, 1683
\bibitem[Lord(1992)]{atran} Lord, S. 1992, NASA-TM-103957
\bibitem[Mastenbrook(1968)]{mas68} Mastenbrook, H. J. 1968, JAtS, 25, 299
\bibitem[Paine(2011)]{am} Paine, S., 2011, SMA memo, 152
\bibitem[Strong et al.(2008)]{str08} Strong, K., Wolff, M. A.,
    Kerzenmacher, T. E. et al. 2008, ACP, 8, 4759
\bibitem[Urban et al.(2004)]{moliere} Urban, J., Baron, P.
    \& Lauti\'e, N. 2004, JQSRT, 83, 529
\bibitem[U.S. Standard Atmosphere(1976)]{us76} U.S. Standard Atmosphere
    1976, NASA-TM-X-74335
\bibitem[Young et al.(2012)]{you12} Young, E. T., Becklin, E. E.,
    De Buizer, J. M., et al. 2012, \apj
\end{thebibliography}
\end{document}